# Novel Phonoreceptive Mechanism of the Cochlear for Low-frequency Sound

W. Yao,[1*] J. Ma,[1] J. Liang,[2] and X. Huang[3]

1. Shanghai Institute of Applied Mathematics and Mechanics, Shanghai 200072, P. R. China.
2. Department of Chemistry, Akron University, Akron, Ohio 44325, USA
3. Department of Otorhinolaryngology, Zhongshan Hospital, Fudan University, Shanghai 200032, P. R. China.

[*]*Corresponding author*, Correspondence and requests for materials should be addressed to W. Yao. (email: wjyao@shu.edu.cn).

## Abstract

Because the cochlear is very small and complex, vibration data of the whole basement membrane are not yet available from existing experiments, To address this question, this work technically adopts the mathematical and biological methods to establish a theoretical analytical model of the Spiral cochlear , combined with medical and modern light source imaging experimental data. In addition, a numerical model of a real human ear is also established. By performing numerous calculations, the results reproduce the known travelling wave vibration of basement membrane. Meanwhile, an exciting finding that revealing a new vibration mode is obtained. More importantly, this newly discovered model intrinsically explain many experimental observations that cannot be explained by travelling wave theory, which solves a long standing various queries to travelling wave vibration among researchers. These results not only complement vibration data that are inaccessible through experiments but also reveal a new hearing mechanism.

## Introduction

The investigation of phonoreceptive mechanism of human cochlear is a major medical problem for mankind. It has received wide attention from many medical and life scientists[1, 2]. Up to now, scientists still have doubts about the vibrating mode of the macrostructure basement membrane. The achievement of auditory function in the cochlear is attributed to fluid-structure coupling motions between lymph and basilar membrane. Revealing the kinetics of basement membrane in the lymph of cochlear is a primary and pivotal question to understand phonoreceptive mechanism of human cochlear [3].

Regarding this crucial question, Békésy proposed his famous travelling wave theory [4] by measuring the motions of BM from fresh temporal bones,. From then on, numerous researchers have proposed mathematical and mechanical models[5-10] to describe the observed vibration behaviour of the basement membrane based on the travelling wave theory. But with the progress of research in this field, many scientists have raised doubts about the travelling wave theory since the end of the last century, doubts originating from the nonviability of travelling wave theory in explaining many experimental discoveries and the low-frequency sound hearing of the cochlear[11-14].

It is worth mentioning that the low-frequency hearing ability is attributed to the apical mechanical behaviour of basement membrane[15-16]. However, it is so difficult to obtain the coupling biological dynamic process between basement membrane and lymph in low-frequency via vivo experiments. Because the apical space is small and the structure is complex, the experimental method for measuring basement membrane motion often perforate Reissner's membrane [17], thus destroying the physiological enviroment of the cochlea and cannot obtain accurate amplitude data. So far, mechanical responses at the apex of the basement membrane are lack, which restrains from deciphering sensory mechanism of cochlea at low-frequency. Moreover, artificial cochlear implant can only improve the hearing in medium or high frequency of the patients with sensorineural deafness, and patients are still deaf at low frequencies.

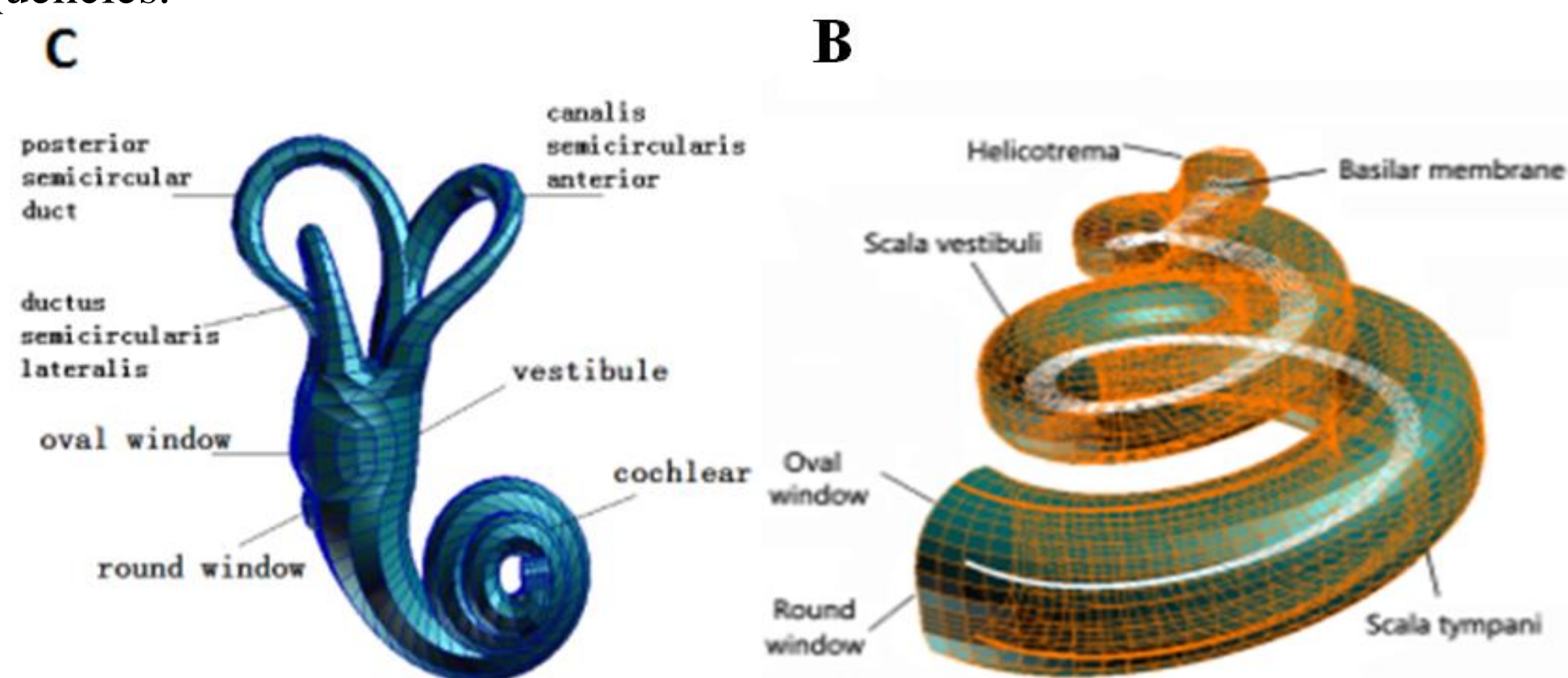

**Figure 1. Numerical model of the human ear**. **(A)** Numerical model of the human inner ear, including the cochlea, vestibule and semicircular canal. The vestibule and semicircular canal are responsible for the human sense of balance, whereas the cochlea is responsible for the sense of hearing. In this paper, we focus on the hearing sensation within the cochlea. **(B)** A 3-D reconstruction of the human cochlea model in a spiral shape within the basement membrane and lamina spiralis ossea, which separates the cochlear into two cavities named the scala vestibuli and scala tympani that are filled with lymph and enclosed by the oval window, round window and surrounding bony shell.

In view of above problems, based upon our decade spanning endeavor in the field of biomechanical investigation of human ear, this work first adopted the Mathematical methods to establish a three-dimensional spiral cochlear fluid-structure mechanical analytical model. Meanwhile, in order to make our research more accurate and credible, a most accurate human ear numerical simulation model was established, based on the most advanced CT imaging technology in Shanghai Synchrotron Radiation Facility. This model was completely in line with morphology of human ear, physiological environment and properties of biological materials. This accurate analytical model and the numerical model characterized by above remarkable features have not been reported in past works.

The presented analytical solutions and numerical simulation results predict the displacement amplitude of basement membrane at low frequencies. A new and exciting discovery is found that there exists a new vibration mode , which is notably distributed on the basement membrane at low frequencies. This finding has not been reported so far. More importantly, the new vibration mode can explain the experimental phenomena which are unexplainable by the travelling wave theory, thus giving an answer to the questions of many scientists about the travelling wave theory for many years.

## Results

### Verification of the presented models

To verify the analytical and numerical models, we calculated the relationship between the magnitude of the basement membrane and the frequency at the position 12 mm from the oval window, which were compared with the measurements[18-19] and, shown in Fig. 2. The results showed good agreement with both measurements with the best frequency of 3000Hz, especially for experimental data 2.

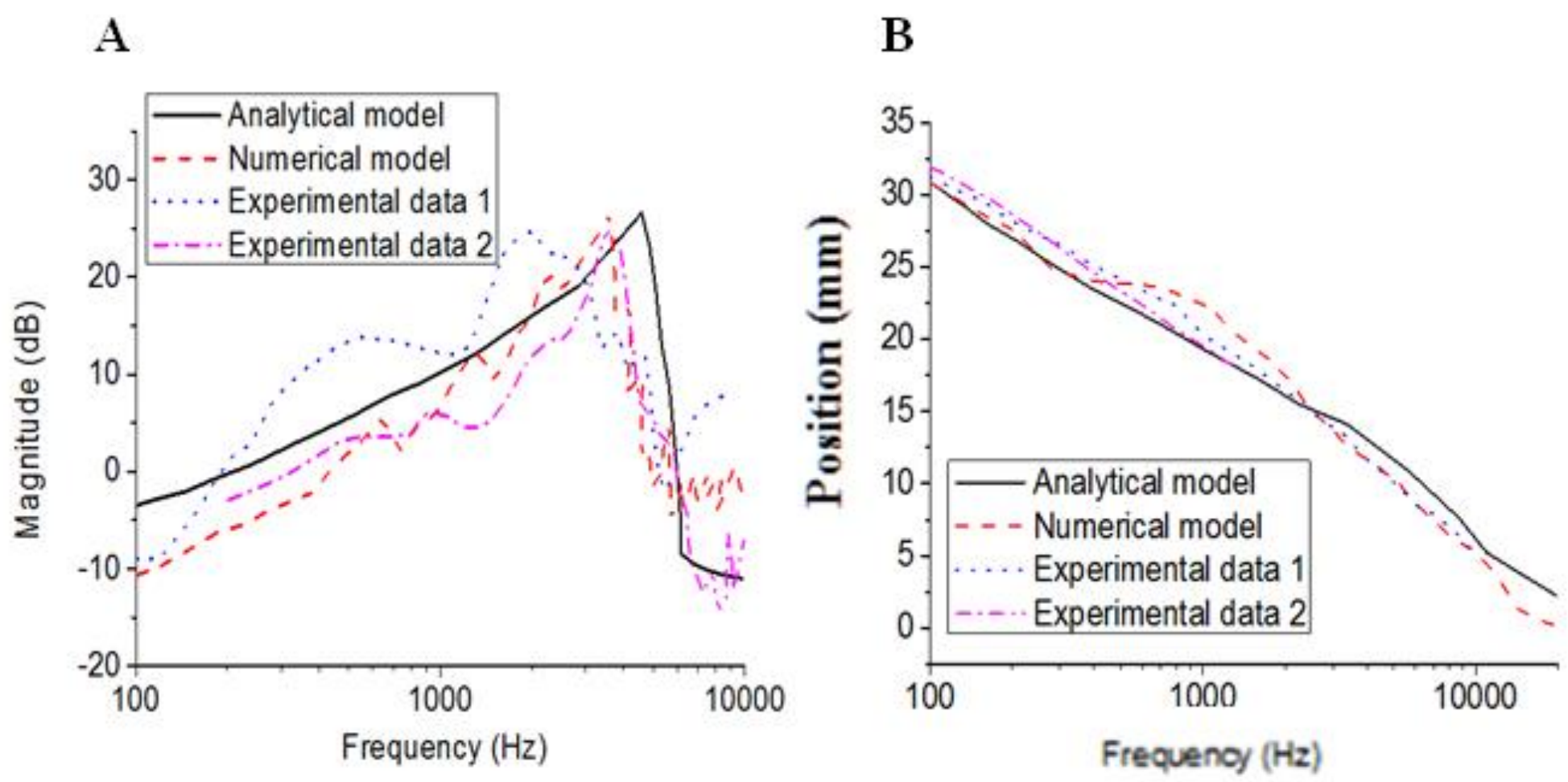

**Figure 2. Comparison of the basement membrane magnitude (observed at x=12 mm from the base) based on the presented model with published experimental data 1 and 2, respectively, by Stenfelt et al.[18] and Gunderson et al.[19] and frequency-position mapping relation** (**A**) Displacement amplitude and (**B**) The frequency-peak mapping curve.

It is worth noting that, in previous numerical models, the cochlear was assumed to be a straight shape and the lymph was assumed to be an incompressible, which is not consistent with the real morphology and biomaterial properties of the cochlear [20-21]. Thus, the previous results were not consistent with the experimental measurements. By comparison, as the presented results shown in Fig. 2, before the best frequency, the curve is not smooth but shows little pits. After the best frequency, the curve declines sharply with increasing frequency until reaching a certain high frequency, and then appears jugged. This finding is consistent with test results that have never been reflected in previous numerical models. This result may have some connection with the new vibration mode, which will be discussed in the remainder of this paper.

The frequency-peak mapping curve of the basement membrane is obtained and compared with the measurements [4, 22]. We find that the simulation curve is consistent with the test curve. In addition, we predict results at very low frequencies that cannot be obtained via the existing measurements.

## basement membrane vibration

Based on the analytical model and the numerical model, the vibration amplitude of the basement membrane at frequencies 200Hz were calculated, as shown in Fig.3. At low frequencies (200 Hz), the basement membrane vibrates with equal amplitude at a regular distance along the spiral longitudinal direction, exhibiting a characteristic the new vibration mode. The new wave forms nodes and anti-nodes, where the amplitudes of the new wave are maximum and minimum, respectively. As the frequency rises, the displacement of the basement membrane increases, and the distance between adjacent and anti-nodes decreases. In addition, all the points between the nodes are in phase, whereas the points on either side of a node are exactly out of phase, undergoing a phase change of $\pi$ radians.

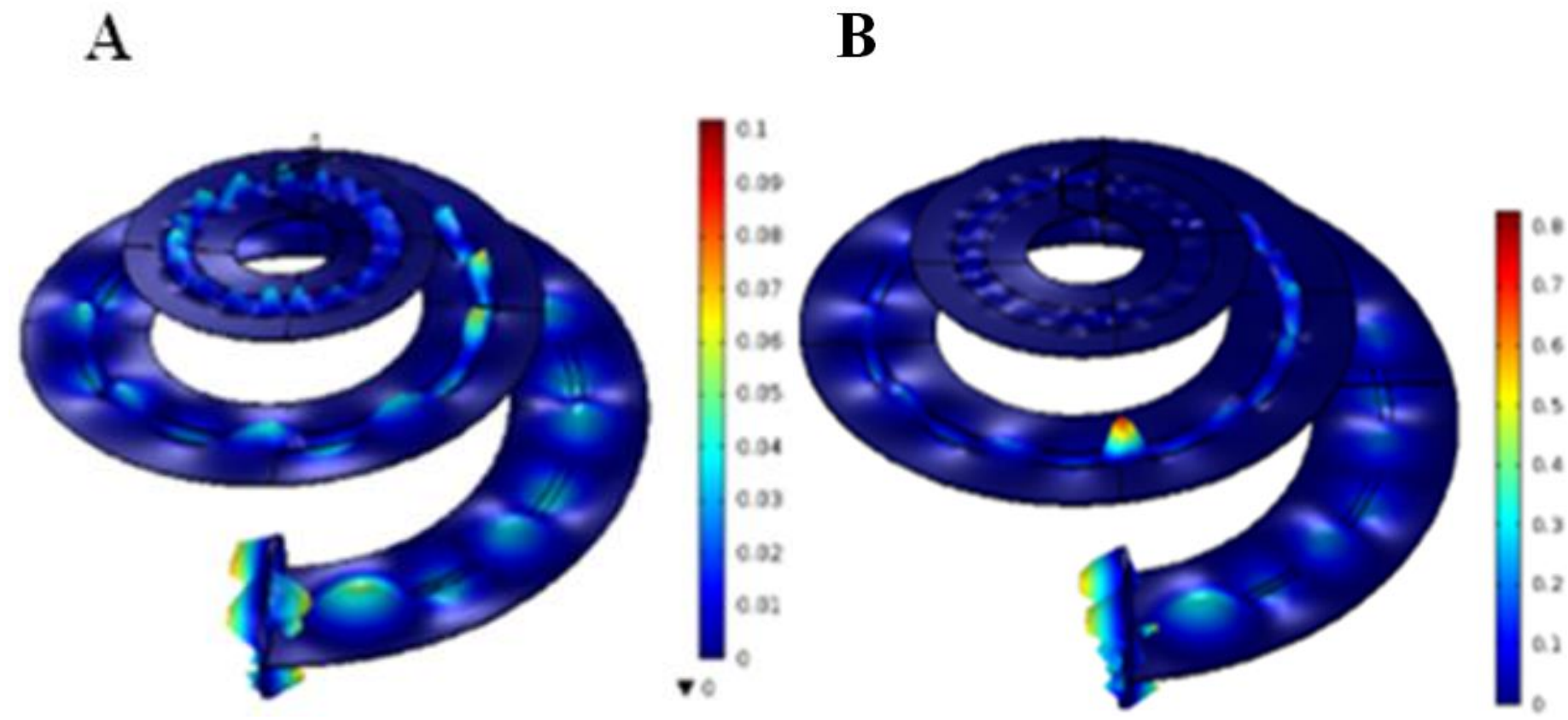

**Figure 3. basement membrane amplitude at low and middle frequencs . (A)** Simulation cloud at 200Hz frequency. **(B)** Simulation cloud at 2500Hz.

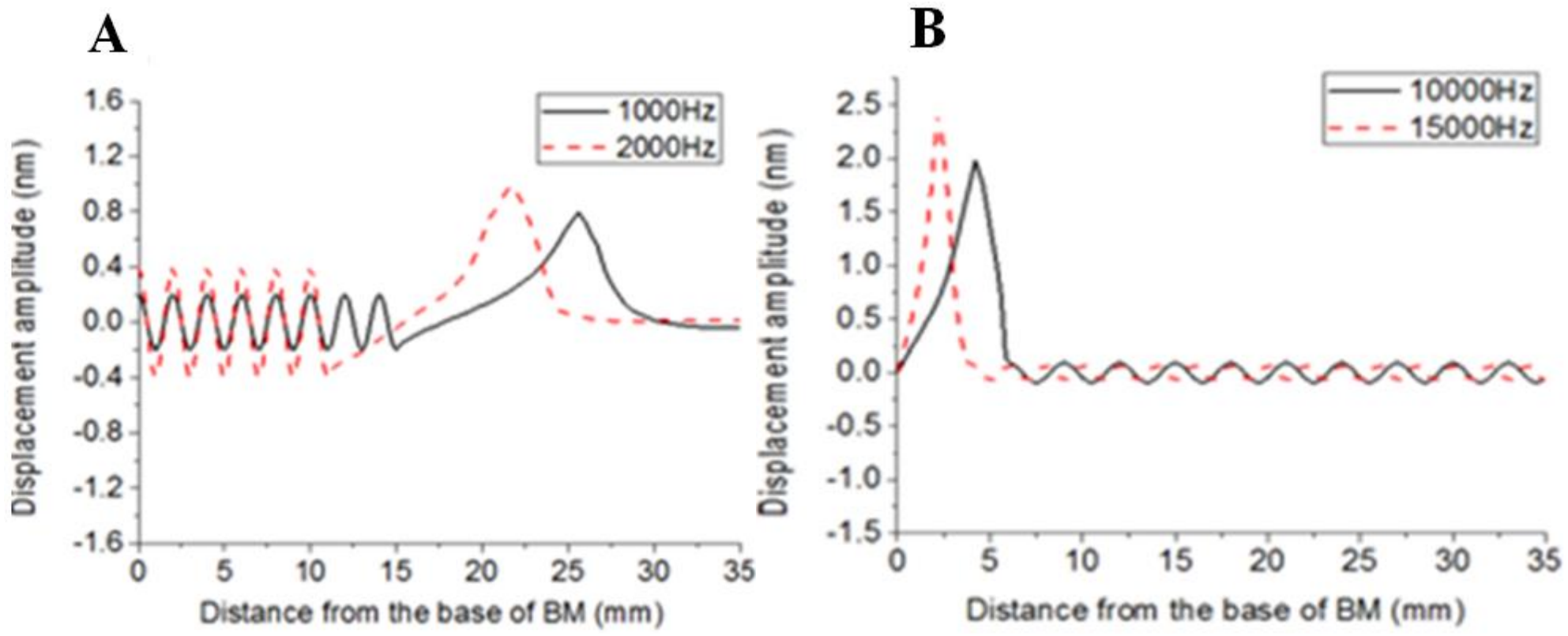

**Figure 4. basement membrane amplitude at middle and high frequencies. (A)** Analytical results at middle frequencies. **(B)** Analytical results at high frequencies

Compared with the analytical solution, at the ends of basement membrane , the new wave pattern has a variable amplitude, which may be attributed to the more accurate geometrical boundaries, material properties and lymph physiological environment of human ear numerical model. The variation of magnitudes of the basement membrane with longitudinal positions at two middle frequencies 1000Hz and 2000Hz are shown in Fig.4. The new wave and travelling wave are concomitant. With the increase of the distance from the cochlear base, there exists a transition of the basement membrane vibration from new wave mode to travelling wave mode. The vibrating amplitude of the new wave pattern is comparatively lower than that of the travelling wave pattern which has a notable single peak. As the frequency rises, the peak of the response curve shifts from the apex to the base.

## Discussion

According to the above results, below 200Hz frequency, the basement membrane mainly vibrates in a new wave form. At frequencies between 1000Hz and 20000Hz, the travelling wave vibration becomes dominant, appearing a peak at the corresponding characteristic frequency. Due to the damping of basement membrane and lymph, after the peak, the travelling wave sharply declines before reaching the apical end. Below 200Hz, the vibration of the basement membrane can reach the apical region. Due to the restraint of top boundary condition of cochlear, the travelling wave reaching the apex will reflect. At the same time, the reflected wave is superimposed with the travelling wave from the bottom, which makes the basement membrane occurring multiple-peak equal amplitude vibration along longitudinal direction.

It is worth noting that, different from the previous numerical models, the material properties of the model in this paper are completely in line with physiological properties of human ear. The lymph is a compressible fluid with low viscosity. In addition, the damping of the basement membrane and lymph are introduced, which can accurately describe the dissipation characteristics of the travelling wave vibration of the basement membran at middle and high frequencies. But the new wave is a non-dissipating wave with arranged nodes and anti-nodes, which is an inherent vibration mode for perceiving low tones.

In fact, since the travelling wave theory was proposed, scientists have discovered that many experimental phenomena cannot be explained by the theory. Kleiner[11] and Richard[12] separately pointed out that at low frequencies, the basement membrane cannot vibrate as described by the travelling wave theory and possibly vibrated in the same amplitude with multiple peaks. Proctor and Zandt experimentally discovered that at low frequencies, it was impossible to find corresponding characteristic positions of single peaks of the basement membrane vibration as depicted by travelling waves [13]. Bell discovered in his experiments that no many phases were accumulated or delayed at characteristic frequencies for vibration of the basement membrane[23]. After experiments, Naidu and Mountain found that the test results of stiffness distribution of the basement membrane could hardly correspond to characteristic frequencies of the membrane[24], so the vibration of basement membrane cannot be explained by the travelling wave theory at some frequencies. Shera proposed that vibration of traveling waves cannot explain the famous otoacoustic emission function of the cochlea [25].

The new wave vibration discovered in this paper could describe the above phenomena that unable to be explained by the travelling wave theory. We discovered that on the basement membrane , there exhibits a new wave which causes the basement membrane vibrate with nodes and anti-nodes at regular distance along the longitudinal direction, appearing multiple peaks with equal amplitude (consistent with the predictions of Kleiner [11], Richard [12], Proctor and Zandt [13]). However, the new wave is a non-propagating wave without phase accumulation from base to the apex (coincide with Bell observations [23]). Compared with the single-peak travelling wave vibration in medium-high frequencies, the new wave vibration in low-frequency is a global and severe vibration occurring entirely on the basement membrane along the longitudinal direction. Therefore, there is no definite relation between the characteristic frequency and correspondent individual position of the basement membrane (consistent with the observations of Naidu and Mountain [24]). The new vibration mode, is accompanied by the generation of reflecting waves (verified Shera's predictions in his experiments [25]).

As mentioned above, the basement membrane vibration data at low-frequency, which were not available through experiments, were obtained in the presented model. Meanwhile, the discovered new vibration mode answers many scientists' doubts about the traveling wave theory and interprets many experimental observations which cannot be explained by the traveling wave theory.

The new vibration mod has explained the coupling mechanical behaviours of low-frequency sounds in cochlear. Not single response was made in certain position on the top of basement membrane whereas vibration within the same amplitude is discovered along the entire basement membrane. This new vibration mode not only enlarge sensation area of basement membrane, but also strengthen abilities of cochlear to hear low-frequency sounds. Such inherent vibration modes at low frequencies are fundamental for cochlear to sense and encode low-frequency sounds. Besides, it possibly affects the active function of cochlear, and thus explain the active phonoreceptive mechanism of the cochlear which is unexplainable by the travelling wave theory.

## Acknowledgments

We would like to thank Zhong Shan Hospital for providing the fresh human right ear specimen and Shanghai Synchrotron Radiation Facility for CT scanning. This work has been supported by three National Natural Science Foundation of China(11072143, 11572186, 11272200) and a China Postdoctoral Science Foundation (2018M632079).

## Competing interests

The authors declare no competing financial interests.